%% file: main-demo.tex
\documentclass[10pt, conference, letterpaper]{IEEEtran}

\IEEEoverridecommandlockouts

\usepackage{cite}
\usepackage{amsmath,amssymb,amsfonts}
\usepackage{algorithmic}
\usepackage{graphicx}
\usepackage{textcomp}
\usepackage{xcolor}
\usepackage{url}
\def\BibTeX{{\rm B\kern-.05em{\sc i\kern-.025em b}\kern-.08em
    T\kern-.1667em\lower.7ex\hbox{E}\kern-.125emX}}
    
\usepackage{fancyhdr}
\begin{document}

\title{Demo: BeGREEN Intelligence Plane for AI-driven Energy Efficient O-RAN management\\
\thanks{This work is supported by the Smart Networks and Services Joint Undertaking (SNS JU) under the European Union’s Horizon Europe research and innovation programme under Grant Agreement No 101097083, BeGREEN project. Views expressed are however those of the author(s) only and do not necessarily reflect those of the European Union or SNS-JU. Neither the European Union nor the granting authority can be held responsible for them.}
}
\author{\IEEEauthorblockN{Miguel Catalan-Cid, David Reiss}
        \IEEEauthorblockA{
            \textit{i2CAT Foundation, Spain}\\
            \{miguel.catalan, david.reiss\}@i2cat.net
        }
\and
\IEEEauthorblockN{German Castellanos}
\IEEEauthorblockA{
            \textit{Accelleran, Belgium}\\
            german.castellanos@accelleran.com 
        }
\and
\IEEEauthorblockN{Joss Armstrong}
\IEEEauthorblockA{
            \textit{Ericsson, Ireland}\\
            joss.armstrong@ericsson.com  
        }
}

\maketitle
\thispagestyle{fancy} 

\begin{abstract}

Cellular networks management is being enhanced by O-RAN architecture and AI/ML solutions, enabling automated intelligent control loops for RAN optimization across various use cases.  Ensuring energy sustainability is crucial to minimizing the impact of mobile networks on global energy consumption. This demo showcases the BeGREEN Intelligence Plane, an AI-driven solution for energy-efficient management of O-RAN networks. The presented workflow focuses on controlling the operational status of emulated cells, highlighting the integration of key components such as the AI Engine and the optimizations achieved through rApps and xApps.
\end{abstract}


\input{Introduction}

\input{Architecture}

\input{Demo}

\bibliographystyle{unsrt}
\bibliography{references.bib}
\end{document}

%% file: Introduction.tex
\section{Introduction}\label{sec:intro}
As emphasized in the IMT-2030 report by ITU \cite{itu}, a key objective for 6G is to minimize network-wide energy consumption. To tackle this challenge, the O-RAN Alliance is developing control mechanisms to manage energy-saving features in multi-vendor environments. The incorporation of Artificial Intelligence and
Machine Learning (AI/ML) will be essential to analyze historical data, adapt proactively to changing network conditions, and drive automated decision-making for energy optimization.

To address these challenges, the SNS BeGREEN project\footnote{\url{https://www.sns-begreen.com/}} proposes an Intelligence Plane, which works as a cross-domain management entity, integrating control and monitoring functions across RAN, Core and Edge domains, and fostering the creation of advanced ML models hosted in its AI Engine component. This paper presents the implementation of the integrated BeGREEN architecture demonstrating one of the targeted use cases, which aims at automating the operational state of network cells, dynamically adapting to forecasted traffic demands.

%% file: Architecture.tex
\section{Architecture}\label{sec:arch}

\begin{figure}[t]
  \centering
  \includegraphics[width=\columnwidth]{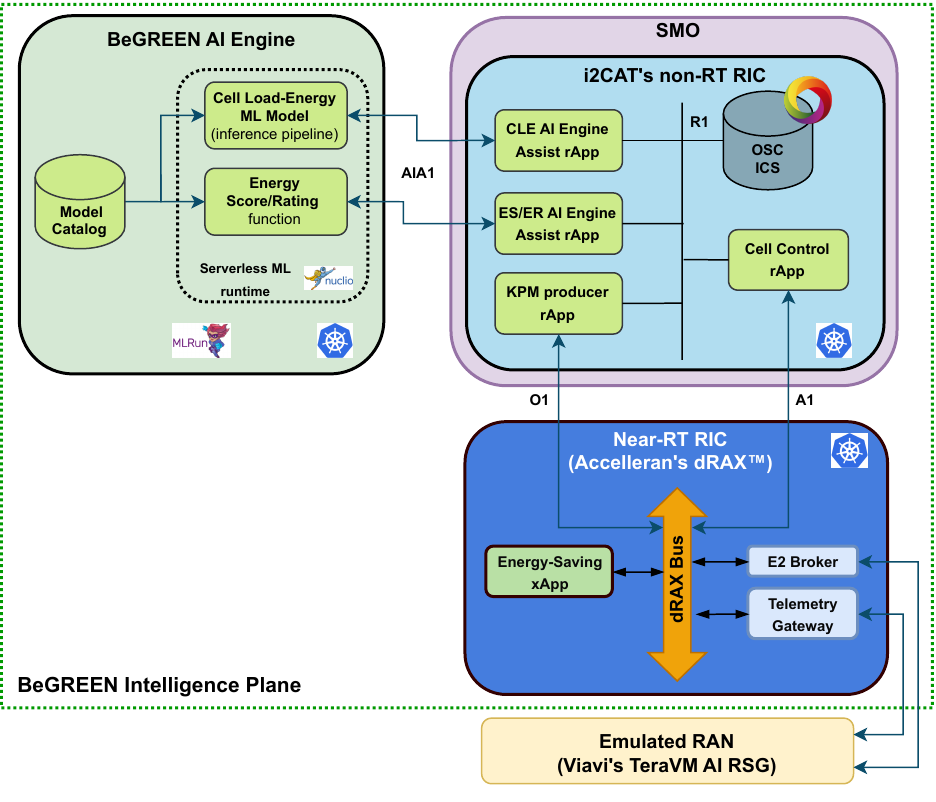}
  \caption{Demo architecture including the BeGREEN Intelligence Plane}
  \label{fig:int_plane}
\end{figure}

The demo showcases the three key components of the BeGREEN Intelligence Plane, as illustrated in Figure \ref{fig:int_plane}. The Non-Real-Time RAN Intelligent Controller (\textbf{Non-RT RIC}), hosts rApps for RAN optimization, utilizing AI/ML models trained on data collected from the emulated E2 nodes. These models are deployed in the \textbf{AI Engine}, which leverages the MLRun\footnote{\url{https://www.mlrun.org/}} and Nuclio\footnote{\url{https://nuclio.io/}} frameworks to enable serverless model inference through AI Engine Assist (AIA) rApps. AIA rApps decouple ML model management from control loops, acting as a proxy between ML models and control rApps. During inference, AIA rApps function as data producers, exposing model outputs via the Data Management and Exposure (DME) service of the R1 interface. In BeGREEN, this interface is implemented using the Information Coordinator Service (ICS) from the O-RAN Software Community \cite{ics}, a data subscription platform that streamlines interactions between data producers and consumers\cite{begreen_d42}.

The AI Engine also provides \textbf{Energy Score} and \textbf{Energy Rating} functions to identify network areas or components requiring energy-saving policies and to assess the benefits of performed optimizations \cite{begreen_d42}. The Energy Score reports an absolute measure of energy efficiency (bits/Joule) based on data volume and energy consumption, while the Energy Rating monitors relative performance by comparing historical data from equivalent network entities, such as cells of the same vendor within the same area. Based on these insights, control rApps determine appropriate A1 energy-saving policies \cite{oran_a1} and communicate them to the Near-RT RIC over the A1 interface. Both control and AIA rApps depend on RAN node data collected and exposed through a Key Performance Measurement (KPM) producer rApp, which connects to the Near-RT RIC via the O1 interface.

The \textbf{Near-RT RIC} handles fast telemetry and control operations for the RAN. It uses an Energy Savings xApp to manage RAN configurations and implement energy-saving policies received from the Non-RT RIC. Integrated into Accelleran's dRAX\textsuperscript{TM} platform\footnote{\url{https://accelleran.com/ran-intelligent-controller/}}, it features a Telemetry Collector that gathers data from both O-RAN and non-O-RAN compliant interfaces. This telemetry is processed into 3GPP-compliant messages, distributed via the dRAX data bus, exposed through the O1 interface, and preprocessed to be visualized in a Grafana dashboard. Additionally, the Near-RT RIC enforces E2-based RAN control and monitors RAN KPIs by interfacing with the emulated RAN environment.

%% file: Demo.tex
\section{Demo description}\label{sec:demo}
This demo showcases the complete workflow involving the components\footnote{The required components are hosted in different premises and interconnected through a VPN.} illustrated in Figure \ref{fig:int_plane}, demonstrating AI-driven management of network cells to improve energy efficiency while maintaining traffic performance. The scenario involves multiple 5G SA cells, with a subset performing as capacity layer cells that can be switched off during low-demand periods. Stationary and mobile UEs are introduced into the scenario to generate downlink traffic. The emulated RAN is implemented using Viavi's TeraVM AI RSG\footnote{https://www.viavisolutions.com/en-us/products/teravm-ai-rsg}. This tool enables the scalable and realistic evaluation of rApps and xApps by emulating O-RAN-compliant and real-world deployments. It also generates 3GPP-compliant KPMs, enabling the collection of data for AI model training and testing. 

Three components are essential to realize the intelligent automated control loop in this scenario, as described below.

\textbf{Cell Load-Energy ML Model}: Supports control rApp decisions on cell operational status. Specifically, it includes a traffic load predictor trained to forecast the expected load for each cell. Managed by its associated AIA rApp, it requires several KPMs as input, including the cell identifier, current load demand, and the number of connected UEs. These inputs are periodically obtained through a subscription to the KPM producer rApp.

\textbf{Cell control rApp}: Implements the control logic to generate A1 energy-saving policies. Once deployed in the non-RT RIC k8s cluster, it creates subscriptions to the AIA and KPM producer rApps to gather the required inputs for decision-making. Based on the predicted load and the Energy Score/Rating of the cells, it identifies candidate cells to be switched on or off. The generated A1 policies adjust the percentage of energy consumption gradually, as defined by O-RAN \cite{oran_a1}, until the cells are fully switched on or off. The Energy Score/Rating is then used as feedback to assess the efficiency of the applied policies.

\textbf{Energy-saving xApp}: Based on the energy-saving policies received, it determines the operational status of the cells and enforces control via the E2 interface. When policies include multiple cells, the xApp must select the required action for each cell to meet the policy target. Additionally, when a cell is switched off, the required handovers are managed by the Smart Handover xApp integrated into the near-RT RIC framework, with a conflict avoidance mechanism ensuring smooth network operation.

\begin{figure}[t]
  \centering
  \includegraphics[width=0.9\columnwidth]{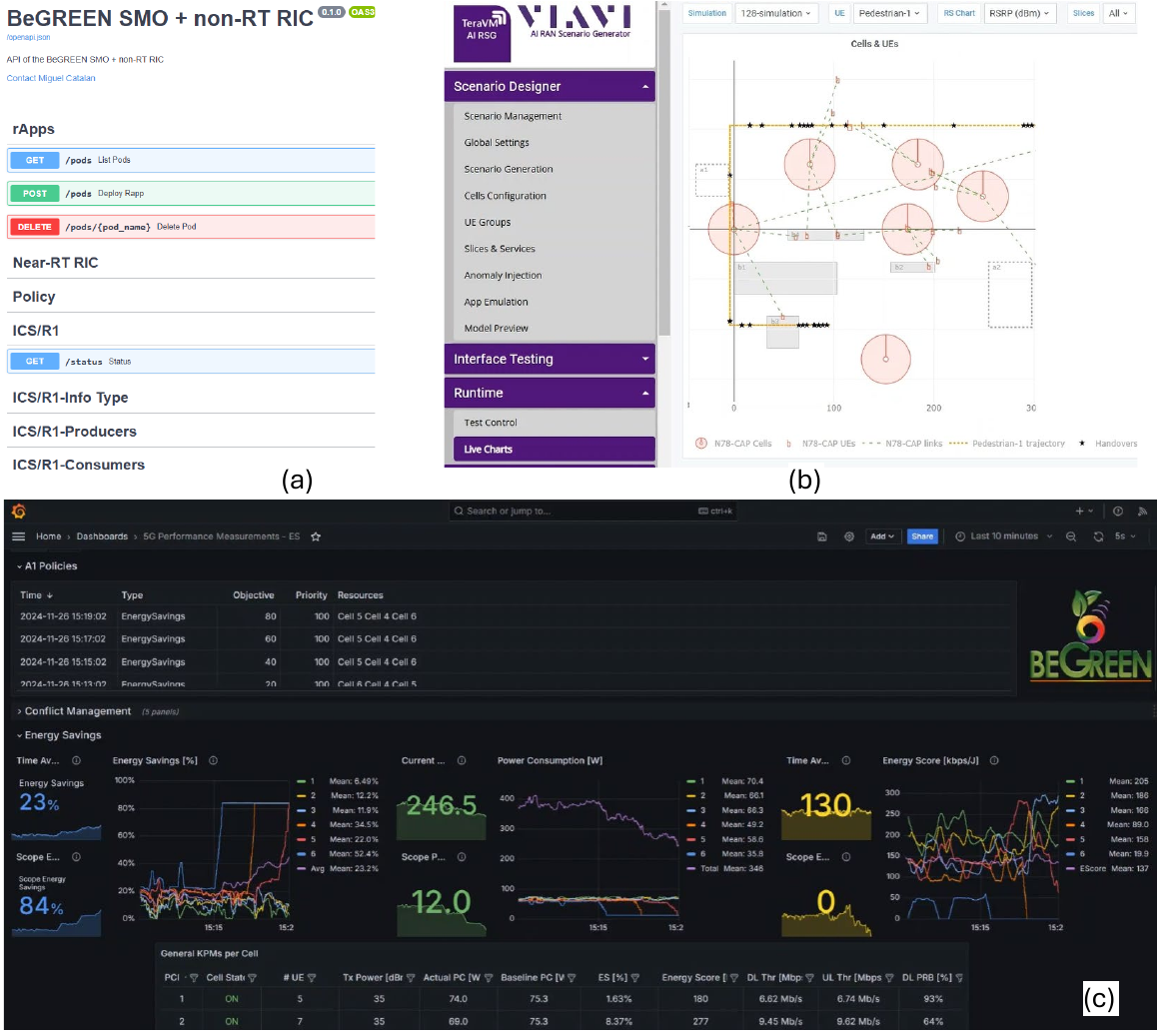}
  \caption{Demo visualization}
  \label{fig:demovis}
\end{figure}

The demo provides insights into the utilization of DME services within the R1 interface, highlighting the data exchange between the involved rApps. This is illustrated through the non-RT RIC Swagger interface, depicted in Figure \ref{fig:demovis}a. Additionally, Viavi's TeraVM AIA RSG dashboard (Figure \ref{fig:demovis}b) presents the scenario details, including the status of cells and UEs. Finally, the dRAX Grafana dashboard (Figure \ref{fig:demovis}c) displays the received A1 policies and the evolution of the KPMs during the demo runtime. Results show energy savings of up to 40\% with no impact to network quality of service. 

